\documentclass[journal,letterpaper,twocolumn,final,10pt]{IEEEtran}

\usepackage{amssymb}
\usepackage{amsthm}
\usepackage{amsmath}
\usepackage{amsfonts}
\usepackage{dsfont}
\usepackage{mathrsfs}
\usepackage{pifont}
\usepackage{amssymb}
\usepackage{verbatim}
\usepackage{upgreek}
\usepackage[final]{graphicx}
\usepackage{epstopdf}
\usepackage{algorithmic}
\usepackage{algorithm}
\usepackage{epsfig}
\usepackage{eucal}
\usepackage{cite}
\usepackage{multirow}
\usepackage{hyperref}
\usepackage{color}

\ifCLASSINFOpdf
\else
   \usepackage[dvips]{graphicx}
\fi
\usepackage{url}

\usepackage{subfigure}

\newcommand{\bbm}{\begin{bmatrix}}
\newcommand{\ebm}{\end{bmatrix}}

\newcommand{\bit}{\begin{itemize}}
\newcommand{\eit}{\end{itemize}}

\newcommand{\ben}{\begin{enumerate}}
\newcommand{\een}{\end{enumerate}}

\newcommand{\bdesc}{\begin{description}}
\newcommand{\edesc}{\end{description}}

\newcommand{\bea}{\begin{array}}
\newcommand{\eea}{\end{array}}

\newcommand{\beqa}{\begin{eqnarray}}
\newcommand{\eeqa}{\end{eqnarray}}

\newcommand{\Comment}[1]{}

\def\C{{\mathds C}}

\def\cC{\mbox{$\CMcal C$}}

\def\cN{\mbox{$\CMcal N$}}

\newcommand{\be}{\begin{equation}}

\newcommand{\ee}{\end{equation}}

\newcommand{\bzero}{{\mbox{\boldmath $0$}}}

\newcommand{\bn}{{\mbox{\boldmath $n$}}}

\newcommand{\bx}{{\mbox{\boldmath $x$}}}

\newcommand{\bz}{{\mbox{\boldmath $z$}}}

\newcommand{\bM}{{\mbox{\boldmath $M$}}}

\newcommand{\bR}{{\mbox{\boldmath $R$}}}
\newcommand{\bS}{{\mbox{\boldmath $S$}}}

\newcommand{\bZ}{{\mbox{\boldmath $Z$}}}

\newcommand{\balpha}{{\mbox{\boldmath $\alpha$}}}

\begin{document}

\title{Adaptive Detection of Dim Maneuvering Targets in Adjacent Range Cells }

\author{Sheng Yan, Pia Addabbo, \IEEEmembership{Senior Member, IEEE}, Chengpeng Hao, \IEEEmembership{Senior Member, IEEE},
and Danilo Orlando, \IEEEmembership{Senior Member, IEEE}
\thanks{This work was in part supported by the National Natural Science Foundation of China under Grant 61971412.}
\thanks{Sheng Yan and Chengpeng Hao are with Institute of Acoustics, Chinese Academy of Sciences, Beijing, China (e-mail: yansheng@mail.ioa.ac.cn;
haochengp@mail.ioa.ac.cn).}
\thanks{Pia Addabbo is with the Universit\`a degli Studi ``Giustino Fortunato,'' 82100 Benevento, Italy (e-mail: p.addabbo@unifortunato.eu).}
\thanks{Danilo Orlando is with the Universit\`a degli Studi ``Niccol\`o Cusano,'' 00166 Roma, Italy (e-mail: danilo.orlando@unicusano.it).}}

\markboth{Journal of \LaTeX\ Class Files, Vol. XX, No. X, September 2020}
{Shell \MakeLowercase{\textit{et al.}}: Bare Demo of IEEEtran.cls for IEEE Journals}
\maketitle

\begin{abstract}
This letter addresses the detection problem of dim maneuvering targets in the presence of range cell migration.
Specifically, it is assumed that the moving target can appear in more than one range cell within the transmitted pulse train.
Then, the Bayesian information criterion and the generalized likelihood ratio test design procedure are jointly exploited to
come up with six adaptive decision schemes capable of estimating the range indices related to the target migration.
The computational complexity of the proposed detectors is also studied and suitably reduced.
Simulation results show the effectiveness of the newly proposed solutions also for a limited set of training data
and in comparison with suitable counterparts.
\end{abstract}

\begin{IEEEkeywords}
Adaptive detection, dim maneuvering targets, range cell migration, radar, sonar, Model Order Selection rules, generalized likelihood ratio test.
\end{IEEEkeywords}

\IEEEpeerreviewmaketitle

\section{Introduction}
\IEEEPARstart{A}{DAPTIVE} detection is a task of primary concern in radar and sonar systems \cite{BOR-Morgan}, \cite{ActiveSonar2017}.
As a matter of fact, in the last decades, a large number of architectures have been developed for the detection of target echoes
competing against noise and clutter interference by means of array of sensors. The common aspect for most
of these contributions is the assumption that the target is point-like and located in the cell under test (CUT) only at a given range.

However, there exist at least three cases where the above assumption may be no longer valid.
Specifically, the first situation concerns high-resolution radars \cite{DistriT} and sonars \cite{RealtimeUnderwater}
which can resolve a target into several scattering centers occupying several consecutive range cells.
In fact, a large amount of detection algorithms for range-spread target can be found in the open
literature (see \cite{DistriT, 9165125,TrainingAssisted, RealtimeUnderwater,4102848} and references therein).

The second case is related to the spillover of target energy between consecutive matched filter samples which
makes a point-like target extended in range \cite{MonopulseRadar} yielding a detection performance degradation when only one sample is processed.
In the seminal paper \cite{LocalizationPoint}, the authors propose a detection architecture that jointly processes adjacent range cells
to take advantage of the spillover limiting the aforementioned degradation.

The third situation arises from the need of increasing the signal-to-interference-plus-noise ratio (SINR) in the case of dim targets
to guarantee reliable detection performance and high-quality target parameter estimates. To this end, radar systems transmit
long bursts of pulses and integrate the corresponding backscattered energy.
However, dim maneuvering targets can move through more than one range cell within the integration time interval \cite{RCMcorr}.
As a consequence, it prevents conventional decision schemes from exploiting all the backscattered energy, since
they are fed by the range bin under test only and, hence, do not account for the target migration to the contiguous range bin.
Therefore, methods to cope with range cell migration (RCM) become of primary importance.
A widely used tool for RCM compensation is the Keystone transform which has been applied in several fields as,
for instance, radar detection \cite{RCMcorr} to mitigate target RCM due to radial velocity and acceleration, synthetic aperture
radar imaging \cite{NewSar}, \cite{RCMSar} where the RCM is caused by linear range walk and range curvature. In \cite{Fastmaneuvering},
an alternative method relying on adjacent correlation function and Lv's
transform is devised to detect the maneuvering targets with radial jerk motion.
More recently, in \cite{DimMovingT}, innovative one-step and two-step detection architectures are conceived
for dim maneuvering targets with and without estimating the slow-time index of the target signal
in the CUT and based upon the generalized information criterion \cite{Stoica1}. Remarkably, such architectures
can overcome conventional detectors as the
generalized adaptive matched filter (GAMF) \cite{GLRT-based} at the price of an
increased computational complexity.

In this letter, we focus on the detection of dim maneuvering targets in the presence of RCM
and further improve the results of \cite{DimMovingT} by devising
innovative robust (with respect to the amount of training samples) architectures.
To this end, we do not consider any
possible phase/amplitude relationships between consecutive pulses and exploit, at the design stage,
the Bayesian information criterion
(BIC) rule \cite{Stoica1}, which is an asymptotic approximation of the optimal maximum a posteriori rule,
to identify the pulse echoes containing target
components over two consecutive range cells.
Then, we conceive two-step architectures (TSA) and one-step architectures (OSA) relying on GLRT-based design criteria, where
GLRT stands for generalized likelihood ratio test. The contributions of the present letter can be summarized as follows:
1) unlike \cite{DimMovingT}, all the samples from two consecutive range cells occupied by the target are processed
to increase the detection performance; 2) the samples free of signal components are exploited for the estimation of
the interference covariance matrix (ICM) lending new architectures a robustness to the training set size;
3) the proposed architectures are designed to avoid a continuous computation of inverse matrices saving computational resources.

The letter is organized as follows: Section II contains the system model and the problem formulation.
In Section III, TSAs and OSAs are devised including suitable modifications of them.
Section IV is devoted to the numerical analysis and discussion. Finally, Section V concludes this letter outlining future research tracks.

\section{System Model and Problem Statement}

\begin{figure}
\centerline{\includegraphics[width=7.2cm]{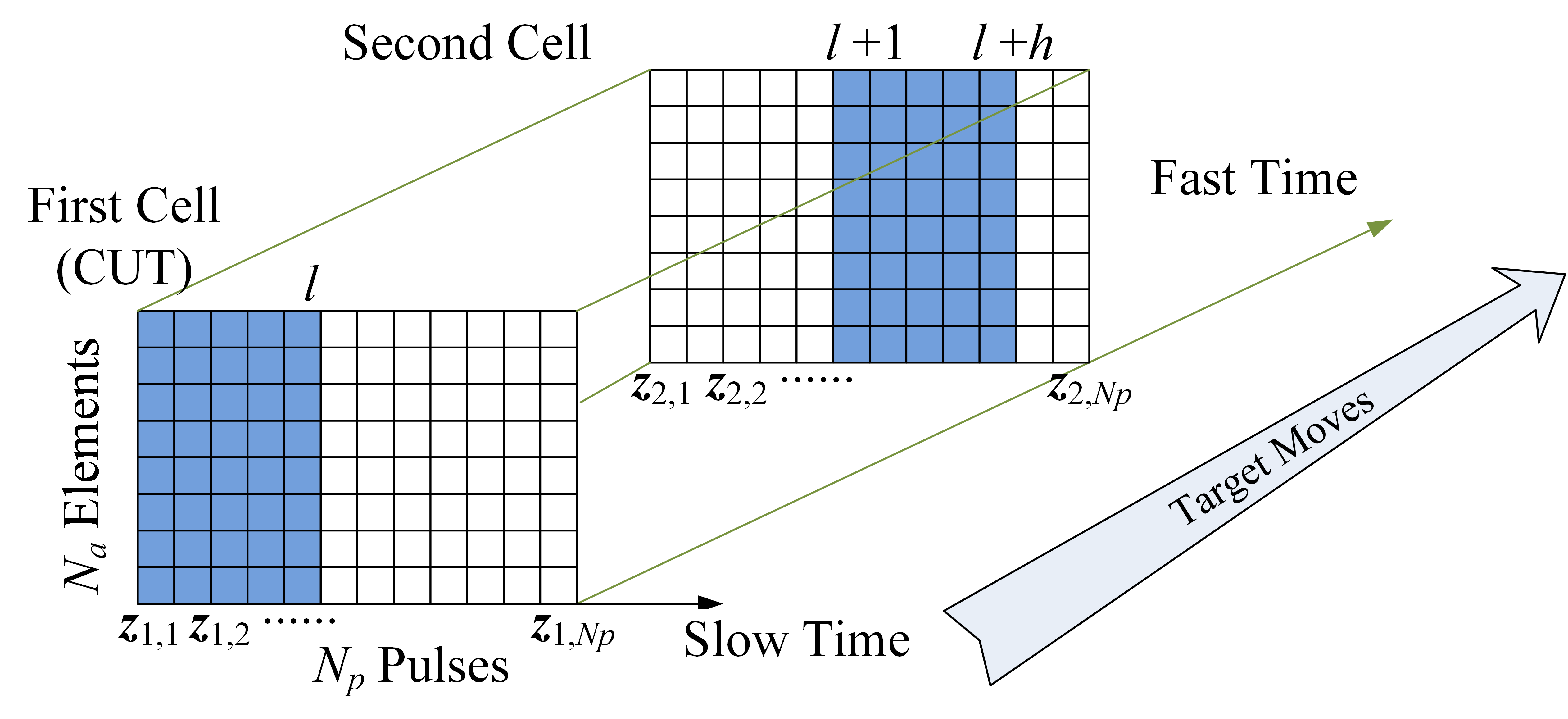}}
\caption{Data matrices in the presence of RCM.}
\label{fig:RCM_figure}
\end{figure}

Let\footnote{\emph{Notation:}
In what follows, vectors and matrices are denoted by boldface lower-case and upper-case letters, respectively.
$(\cdot)^\dag$,  $(\cdot)^T$, and $\det(\cdot)$ denote the complex conjugate transpose, the transpose, and the determinant,
respectively, of the matrix argument, whereas the imaginary unit is $j$.
As to the numerical sets, $\mathbb{C}$ is the set of complex numbers, and $\mathbb{C}^{N\times M}$
is the Euclidean space of $(N\times M)$-dimensional complex matrices (or vectors if $M=1$).
The symbol $\Re\{\cdot\}$ indicates the real part of a complex number, and the symbol $\triangleq$ denotes a definition.
The $i$th entry of a vector $\bx$ is indicated by $\bx(i)$ and $\bzero$ denotes the null vector whose size depends on the the context.
Finally, we write $\boldsymbol{x}\sim\cC\cN_N(\boldsymbol{m},\boldsymbol{M})$ if $\boldsymbol{x}$ is a complex
circular $N$-dimensional normal vector with mean $\boldsymbol{m}$ and positive definite covariance matrix $\boldsymbol{M}$.}
us consider a (radar or sonar) system equipped with a linear array of $N_a$ identical and uniformly distributed sensors
(the inter-element spacing $d$ is half of the operating wavelength, $\lambda$ say, to avoid spatial aliasing).
Moreover, denote by $N_p$ pulses belonging to the transmitted pulse train.
Then, for a point-like target, the signal received by the $m$th antenna element can be written as \cite{DimMovingT}
\begin{eqnarray}
x_m(t) = \Re\bigg\{\alpha \sum_{n=0}^{N_p-1}p\big(t-nT-\tau_0+\frac{2v_t}{c}nT\big)
\nonumber\\
\times e^{j2\pi(f_c+f_d)t}e^{j2\pi(m-1)\nu_s}\bigg\},
\end{eqnarray}
where $\alpha\in\C$ accounts for target and channel effects, $T>0$ is the pulse repetition
time (PRT), $p(t)$ is an unit-energy pulse waveform, $\tau_0$ is the round-trip delay of the target,
$v_t$ is the target radial velocity, $c$ is the waveform velocity of propagation,
$f_c$ is the carrier frequency, $f_d$ is the target Doppler frequency,
and $\nu_s = \frac{d\cos\psi}{\lambda}$ is the spatial frequency with $\psi$ the nominal target angle of arrival (AOA).
After matched filtering and digital sampling,
for the $q$th range cell (fast time) which is the target location at $t = 0$ we obtain the data sequence
as \cite{LocalizationPoint}
\begin{align}
&y_m(q,g) = \alpha e^{j2\pi f_d T(g-1)} \mathcal{X}_p\big(\tau_0-(q-1)T_p\nonumber
\\
&+(g-1)\frac{2v_t}{c}T, f_d\big) e^{j2\pi(m-1)\nu_s}
\triangleq \alpha(q,g) e^{j2\pi(m-1)\nu_s},
\label{eqn:sig_model_RCM}
\end{align}
where $g\in \Omega_p=\{1,\ldots, N_p\}$ indexes the slow time, $(g-1)\frac{2v_t}{c}T$ accounts
for the range migration, $\mathcal{X}_p(\cdot,\cdot)$
is the ambiguity function of $p(t)$, $T_p$ is the one-sided mainlobe width of the zero-Doppler cut of $\mathcal{X}_p$,
and $\alpha(q,g) = \alpha e^{j2\pi f_d T(g-1)} \mathcal{X}_p\big(\tau_0-(q-1)T_p+(g-1)\frac{2v_t}{c}T, f_d\big)$.
Equation \eqref{eqn:sig_model_RCM} highlights that, in the case of maneuvering targets and for large values of $g$ (and, hence, of $N_p$),
target response (ambiguity function value) associated with the $q$th range bin can decrease to zero implying that
the mainlobe of the ambiguity function has migrated to the next contiguous range bin, namely the RCM has occurred.
In what follows, for simplicity and without loss of generality, we set $q = 1$ and, hence, the next contiguous range bin is indexed by
$q=2$. Now, let us define by $\boldsymbol{Z}_{i} = [\boldsymbol{z}_{i,1},\boldsymbol{z}_{i,2},\ldots,\boldsymbol{z}_{i,N_p}]
\in\C^{N_a\times N_p}$, $i=1,2$,
the data matrix corresponding to the $i$th range bin whose columns contain the returns from the $N_a$ spatial channels. Then,
if we assume that the $\bar{g}$th echo from range bin $1$ contains target components and that the same echo from range bin $2$ is representative
of interference only, we can write $\boldsymbol{z}_{1,\bar{g}} = [y_1(1,\bar{g}),\ldots,y_{N_a}(1,\bar{g})]^T+\boldsymbol{n}_{1,\bar{g}}\,
\triangleq \alpha(1,\bar{g}) \boldsymbol{v}+\boldsymbol{n}_{1,\bar{g}}$
and $\boldsymbol{z}_{2,\bar{g}} = \bn_{2,\bar{g}}$,
where $\boldsymbol{v}=[1, e^{j2\pi \nu_s},\ldots,e^{j2\pi(N_a-1)\nu_s}]^T \in \mathbb{C}^{N_a \times 1}$ is the nominal spatial steering
vector depending on $\psi$ and the $\boldsymbol{n}_{i,g}$s are the interference components. When the RCM occurs for some pulse index $\tilde{g}$,
previous situation changes, namely $\bz_{1,\tilde{g}}=\bn_{1,\tilde{g}}$ is representative of interference
only whereas $\bz_{2,\tilde{g}}=\alpha(2,\tilde{g}) \boldsymbol{v}+\bn_{2,\tilde{g}}$ also contains target components.
A pictorial description of the RCM is shown in Fig. \ref{fig:RCM_figure}, where the blue squares denote data
with target components and white squares denote data free of useful signal echoes.
In the first $l$ pulses, the target is in the first cell, then, it moves to the second range cell.

Therefore, in order to account for possible range migration, it is reasonable to process the returns
associated with (at least) two consecutive range cells.
Summarizing, the detection problem at hand can be formulated in terms of
the following multiple hypothesis test
\begin{eqnarray}
\left\{
\begin{array}{l}
H_{l,h} : \left\{
\begin{array}{lr}
\boldsymbol{z}_{1,l+1} = \boldsymbol{n}_{1,l+1},\ldots, \boldsymbol{z}_{1,N_p} = \boldsymbol{n}_{1,N_p}, & \\
\boldsymbol{z}_{2,1} = \boldsymbol{n}_{2,1},\ldots, \boldsymbol{z}_{2,l} = \boldsymbol{n}_{2,l}, & \\
\boldsymbol{z}_{2,l+h+1} = \boldsymbol{n}_{2,l+h+1},\ldots, & \\ \boldsymbol{z}_{2,N_p} = \boldsymbol{n}_{2,N_p}, & \\
\boldsymbol{z}_{1,1} = \alpha(1,1)\boldsymbol{v}+\boldsymbol{n}_{1,1},\ldots, & \\
\boldsymbol{z}_{1,l} = \alpha(1,l)\boldsymbol{v}+\boldsymbol{n}_{1,l}, & \\
\boldsymbol{z}_{2,l+1} = \alpha(2,l+1)\boldsymbol{v}+\boldsymbol{n}_{2,l+1},\ldots,\\ \boldsymbol{z}_{2,l+h} = \alpha(2,l+h) \boldsymbol{v}+\boldsymbol{n}_{2,l+h}, & \\
\boldsymbol{r}_{k} = \boldsymbol{m}_{k},k = 1, \ldots, K, & \\
\end{array}
\right.
\\
H_0 \,\,\,: \left\{
\begin{array}{ll}
\boldsymbol{z}_{1,1} = \boldsymbol{n}_{1,1},\ldots, \boldsymbol{z}_{1,N_p} = \boldsymbol{n}_{1,N_p}, & \\
\boldsymbol{z}_{2,1} = \boldsymbol{n}_{2,1},\ldots, \boldsymbol{z}_{2,N_p} = \boldsymbol{n}_{2,N_p}, & \\
\boldsymbol{r}_{k} = \boldsymbol{m}_{k},k = 1, \ldots, K, & \\
\end{array}
\right.
\end{array}
\right.
\noindent
\label{eqn:det_prob_RCM}
\end{eqnarray}
where $\boldsymbol{R}=[\boldsymbol{r}_{1} \ldots \boldsymbol{r}_{K}]$ are the training data, $ 1\le l \le N_p $, $ 0\le h \le N_p -l$
are unknown integers indexing which vectors contain target components,
$\boldsymbol{n}_{1,i},\boldsymbol{n}_{2,i},\boldsymbol{m}_{i} \sim \cC\cN_{N_a}(\boldsymbol{0},\boldsymbol{M})$
are statistically independent interference vectors. As for $\alpha(1,i), i = 1, \ldots, l$ and $\alpha(2,i), i = l+1, \ldots, l+h$, they
are modeled according to the Swerling II model \cite{Richards}.
Finally, note that when $H_{l,h}$ is declared, the nominal target AOA $\psi$ can be used as a preliminary estimate of the
actual target AOA.

\section{Design Issues}
In this section, we devise two classes of architectures for problem \eqref{eqn:det_prob_RCM}.
The first class pursues a natural approach which consists in estimating the pulse indices corresponding to
the range transition (Subsection III.A) and then in applying decision schemes based upon such estimates
(Subsection III.B). It follows that such architectures consist of two stages (TSA): the first stage solves the RCM problem
whereas the second stage is responsible for target detection. The second approach (OSA) jointly performs the
above operations using a penalized GLRT-based
decision scheme \cite{VanTrees4} (Subsection III.C). Even though from a conceptual point of view
these approaches share the same operations, from an operating point of view they can lead to different performance as shown in Section IV.

\subsection{First Stage of TSA: RCM Estimation}
The preliminary stage of the TSAs is aimed at estimating parameters $l$ and $h$ using two BIC-based selection rules.
More precisely, the first rule is devised according to the two-step design paradigm that consists in applying well-established
design criteria assuming that some parameters are known (first step) and then replacing them with suitable estimates (second step).
An example of this paradigm is provided by \cite{kelly1992AMF} in the context of adaptive radar detection. Thus,
following this line of reasoning, we first assume that $\bM$ is known and derive the BIC rule as follows
\begin{eqnarray}
\min_{\substack{l \in\Omega_p \\ h:  \, l+h \le N_p}} \big\{-2\ln f_{l,h}(\boldsymbol{Z};\hat{\boldsymbol{\alpha}}_{l,h,\boldsymbol{M}},\boldsymbol{M})+p_1(l,h) \big\},
\end{eqnarray}
where $\bZ = [\boldsymbol{Z}_1, \boldsymbol{Z}_2]\in\C^{N_a\times 2N_p}$,
$\hat{\boldsymbol{\alpha}}_{l,h,\boldsymbol{M}} \triangleq [
\frac{\boldsymbol{v}^\dagger \boldsymbol{M}^{-1} \boldsymbol{z}_{1,1}}{\boldsymbol{v}^\dagger \boldsymbol{M}^{-1} \boldsymbol{v}} \ldots$ $
\frac{\boldsymbol{v}^\dagger \boldsymbol{M}^{-1} \boldsymbol{z}_{1,l}}{\boldsymbol{v}^\dagger \boldsymbol{M}^{-1} \boldsymbol{v}}
\frac{\boldsymbol{v}^\dagger \boldsymbol{M}^{-1} \boldsymbol{z}_{2,l+1}}{\boldsymbol{v}^\dagger \boldsymbol{M}^{-1} \boldsymbol{v}} \ldots
\frac{\boldsymbol{v}^\dagger \boldsymbol{M}^{-1} \boldsymbol{z}_{2,l+h}}{\boldsymbol{v}^\dagger \boldsymbol{M}^{-1} \boldsymbol{v}}]^T
\in\C^{(l+h)\times 1}$
is the maximum likelihood estimate (MLE) of $\boldsymbol{\alpha}\!=\![\alpha(1,1) \,\ldots\, \alpha(1,l) \,\, \alpha(2,l+1) \,\ldots \, \alpha(2,l+h)]^T$ for known $\bM$ \cite{GLRT-based},
$f_{l,h}(\boldsymbol{Z};\boldsymbol{\alpha},\boldsymbol{M})$ %is the probability density function (PDF) expression
is the probability density function (PDF) of $\boldsymbol{Z}$ under $H_{l,h}$,
and $p_1(l,h)=2(l+h)\ln(4N_aN_p)$ is the penalty term accounting for the number of unknown parameters ($\balpha$)
and the volume of data.
Finally, replacing $\bM$ with $\boldsymbol{S}/K=\boldsymbol{R}\boldsymbol{R}^\dagger/K$ to achieve adaptivity and
neglecting the irrelevant constants, the final optimization problem is
\be
\min_{\substack{l \in\Omega_p \\ h:  \, l+h \le N_p}} \Bigg\{-2K
\Lambda_{l,h}(\bZ,\bS)
+ p_1(l,h) \Bigg\},
\label{eqn:BIC_2S}
\ee
where $\Lambda_{l,h}(\bZ,\bS)=\sum_{i=1}^l{\frac{\mid \boldsymbol{z}_{1,i}^\dagger \boldsymbol{S}^{-1} \boldsymbol{v} \mid ^2}
{\boldsymbol{v}^\dagger \boldsymbol{S}^{-1} \boldsymbol{v}}}+
\sum_{w=l+1}^{l+h}{ \frac {\mid \boldsymbol{z}_{2,w}^\dagger \boldsymbol{S}^{-1} \boldsymbol{v} \mid ^2}
{\boldsymbol{v}^\dagger \boldsymbol{S}^{-1} \boldsymbol{v}}}$.

The second selection rule consists in applying the BIC criterion over $\boldsymbol{Z}$ and $\boldsymbol{R}$ to obtain
\begin{eqnarray}
\min_{\substack{l \in\Omega_p \\ h:  \, l+h \le N_p}} \Big\{-2\ln \lbrack
f (\boldsymbol{R};\hat{\boldsymbol{M}}_{l,h}) f_{l,h}(\boldsymbol{Z};
\hat{\boldsymbol{\alpha}}_{l,h,\boldsymbol{S}_{l,h}},\hat{\boldsymbol{M}}_{l,h}) \rbrack
\nonumber\\
+ p_2(l,h) \Big\},
\label{eqn:BIC_1S_p}
\end{eqnarray}
where
$\boldsymbol{S}_{l,h}\!\!\!\!=\!\!\!\!\bR\bR^\dagger + \sum_{i=l+1}^{N_p} \!\!{\bz_{1,i}\bz_{1,i}^\dagger}
+ \sum_{w=1}^{l} \!\!{\bz_{2,w}\bz_{2,w}^\dagger} $ $+ \sum_{b=l+h+1}^{N_p} {\bz_{2,b}\bz_{2,b}^\dagger}$,
$
\hat{\boldsymbol{M}}_{l,h} = {\boldsymbol{S}_{l,h}}/{(2N_p+K)} +
{\sum_{i=1}^l{(\boldsymbol{z}_{1,i}-\hat{\balpha}_{l,h,\boldsymbol{S}_{l,h}}(i) \boldsymbol{v})(\boldsymbol{z}_{1,i}
-\hat{\balpha}_{l,h,\boldsymbol{S}_{l,h}}(i) \boldsymbol{v})^\dagger} }/{(2N_p+K)}
+{\sum_{w=l+1}^{l+h}{(\boldsymbol{z}_{2,w}-\hat{\balpha}_{l,h,\boldsymbol{S}_{l,h}}(w) \boldsymbol{v})(\boldsymbol{z}_{2,w}
-\hat{\balpha}_{l,h,\boldsymbol{S}_{l,h}}(w) \boldsymbol{v})^\dagger} }/{(2N_p+K)}
$
is the MLE of $\boldsymbol{M}$ based upon $\boldsymbol{Z}$ and $\boldsymbol{R}$ under $H_{l,h}$,
$f(\boldsymbol{R};{\hat{\bM}_{l,h}})$
is the PDF of $\boldsymbol{R}$ computed at $\hat{\bM}_{l,h}$,
and $p_2(l,h)=(2l+2h+N_a^2)\ln(4N_aN_p+2N_aK)$ is the penalty term.
It is possible to show that (6) is equivalent to
\begin{eqnarray}
\min_{\substack{l \in\Omega_p \\ h:  \, l+h \le N_p}} \big\{(4N_p+2K)\ln\det(\hat{\boldsymbol{M}}_{l,h} )+p_2(l,h)\big\}.
\label{eqn:BIC_1S}
\end{eqnarray}
Notice that the above equation requires the computation of $\hat{\balpha}_{l,h,\boldsymbol{S}_{l,h}}$ and, hence, the
inversion of $\boldsymbol{S}_{l,h}$ for each admissible pair $(l,h)$. To reduce the computational load of \eqref{eqn:BIC_1S}, we
replace $\bS_{l,h}$ with $\bS$, which does not require to be updated.
The reduced-complexity BIC rule is given by
\begin{eqnarray}
\min_{\substack{l \in\Omega_p \\ h:  \, l+h \le N_p}} \{ (4N_p+2K)\ln\det(\hat{\boldsymbol{M}}'_{l,h}) + p_2(l,h)\},
\label{eqn:BIC_1S_LCC}
\end{eqnarray}
where $\hat{\boldsymbol{M}}'_{l,h}=\frac{\boldsymbol{S}_{l,h}}{2N_p+K} + \frac{\sum_{i=1}^l{(\boldsymbol{z}_{1,i}
-\hat{\bf \alpha}_{l,h,\boldsymbol{S}}(i) \boldsymbol{v})(\boldsymbol{z}_{1,i}
-\hat{\bf \alpha}_{l,h,\boldsymbol{S}}(i) \boldsymbol{v})^\dagger} }{2N_p+K}$
$+\frac{\sum_{w=l+1}^{l+h}{(\boldsymbol{z}_{2,w}-\hat{\bf \alpha}_{l,h,\boldsymbol{S}}(w) \boldsymbol{v})(\boldsymbol{z}_{2,w}
-\hat{\bf \alpha}_{l,h,\boldsymbol{S}}(w) \boldsymbol{v})^\dagger} }{2N_p+K}$.
It is worth noticing that the price to be paid for the reduced computational load is
a performance degradation especially when training data are limited as shown in Section IV.

\subsection{TSA Architectures}
The second (detection) stage of TSAs exploits the estimates of $l$ and $h$, denoted by $\hat{l}$ and $\hat{h}$, respectively,
provided by the first stage and the following GAMF-like \cite{GLRT-based} decision rule
\be
\Lambda_{\hat{l},\hat{h}}(\bZ,\bS)\overset{H_{\hat{l},\hat{h}}}{\underset{H_0}{\gtrless}}\eta,
\label{eqn:GAMF_1}
\ee
where $\eta$ is the threshold\footnote{Hereafter, we denote by $\eta$ the generic detection threshold.}
set according to the value of the probability of false alarm ($P_{fa}$) and $\Lambda_{{l},{h}}(\bZ,\bS)$
has been defined after \eqref{eqn:BIC_2S}.
Thus, we can obtain two architectures by cascading \eqref{eqn:GAMF_1} with \eqref{eqn:BIC_2S} (TSA-1) and
\eqref{eqn:GAMF_1} with \eqref{eqn:BIC_1S} (TSA-2).
In addition, the left-hand side of \eqref{eqn:GAMF_1} can be suitably modified to make it less sensitive to the amount of secondary data by
replacing $\bS$ with $\bS_{\hat{l},\hat{h}}$ (see the definition after \eqref{eqn:BIC_1S_p}),
which exploits additional data drawn from those
associated to the range cells under test. Therefore, the modified decision rule is
\be
\Lambda_{\hat{l},\hat{h}}(\bZ,\bS_{\hat{l},\hat{h}})\overset{H_{\hat{l},\hat{h}}}{\underset{H_0}{\gtrless}}\eta.
\ee
The above decision rule can be coupled with \eqref{eqn:BIC_2S} and \eqref{eqn:BIC_1S_LCC} to obtain the modified TSA-1 (M-TSA-1) and
modified TSA-2 (M-TSA-2), respectively.

\subsection{OSA Architectures}
The one-stage detection architectures rely on a ``penalized generalized likelihood
ratio test'' \cite{VanTrees4}, whose penalty term is borrowed from BIC rule, and
jointly perform target detection and RCM estimation without intermediate steps.
Again, we develop two OSAs that differ in the way secondary data are incorporated into the decision statistic.
This first architecture (OSA-1) relies on the GAMF \cite{GLRT-based} and is given by
\be
\max_{\substack{l \in\Omega_p \\ h:  \, l+h \le N_p}}
\biggl\{
\Lambda_{l,h}(\bZ,\bS)
-\frac{p_1(l,h)}{2K}\biggr\} \overset{H_{\hat{l},\hat{h}}}{\underset{H_0}{\gtrless}}\eta.
\ee
The second architecture (OSA-2) is obtained by applying the logarithm of the GLRT over both primary and secondary data, namely
\begin{eqnarray}
\max_{\substack{l \in\Omega_p \\ h:  \, l+h \le N_p}} \biggl\{\ln \lbrack f (\boldsymbol{R};\hat{\boldsymbol{M}}_{l,h})
f_{l,h}(\boldsymbol{Z};\hat{\boldsymbol{\alpha}}_l,\hat{\boldsymbol{M}}_{l,h}) \rbrack -\frac{p_2(l,h)}{2} \biggr\}
\nonumber\\
-\max_{\bM} \biggl\{\ln \lbrack f (\boldsymbol{R};\boldsymbol{M}) f_0(\boldsymbol{Z};\boldsymbol{M}) \rbrack  \biggr\}
\overset{H_{\hat{l},\hat{h}}}{\underset{H_0}{\gtrless}}\eta,
\label{eqn:OSA_2_p}
\end{eqnarray}
where $f_0(\boldsymbol{Z};\boldsymbol{M})$
is the PDF of $\boldsymbol{Z}$ under $H_{0}$. It is possible to show that \eqref{eqn:OSA_2_p} can be recast as
\be
\ln \det \biggl( \! \frac{\boldsymbol{S} \!+\! \bZ\bZ^\dag}{2N_p\!+\!K} \! \biggr)
+\!\!\!\!\!
\max_{\substack{l \in\Omega_p \\ h:  \, l+h \le N_p}}
\!\!\!\!\!
\biggl(\! \ln \frac{1} {\det(\hat{\boldsymbol{M}}_{l,h} )}-\frac{p_2(l,h)}{(4N_p+2K)}\!\biggr)
\!\!\!
\overset{H_{\hat{l},\hat{h}}}{\underset{H_0}{\gtrless}}
\!\!\!
\eta.
\ee

\section{Performance Assessment}
In this section, we investigate the behavior of the proposed architectures in terms of
probability of detection ($P_d$), computational complexity, and mean value of misclassified pulses
(MVMP) defined as the sum of the number of pulses containing target components but classified as noise and the number
of noise-only pulses classified as target (this metric is estimated only for the TSAs since OSAs inherit the selection capabilities of BIC).
The competitors are the likelihood ratio test assuming perfect knowledge of
$l$, $h$, and $\bM$ (clairvoyant detector), the GAMF and the generalized adaptive subspace detector (GASD) \cite{GLRT-based}
both over data from two range cells, and
the best detector of \cite{DimMovingT} defined by (10) and (11)
(2S-GIC) and fed by data from the first range bin. Notice that the clairvoyant detector represents an upper bound for the performance.

\begin{figure}
\centering
\subfigure[$P_d$ versus SINR with RCM.]{
\includegraphics[width=0.48\linewidth, height=3.3cm]{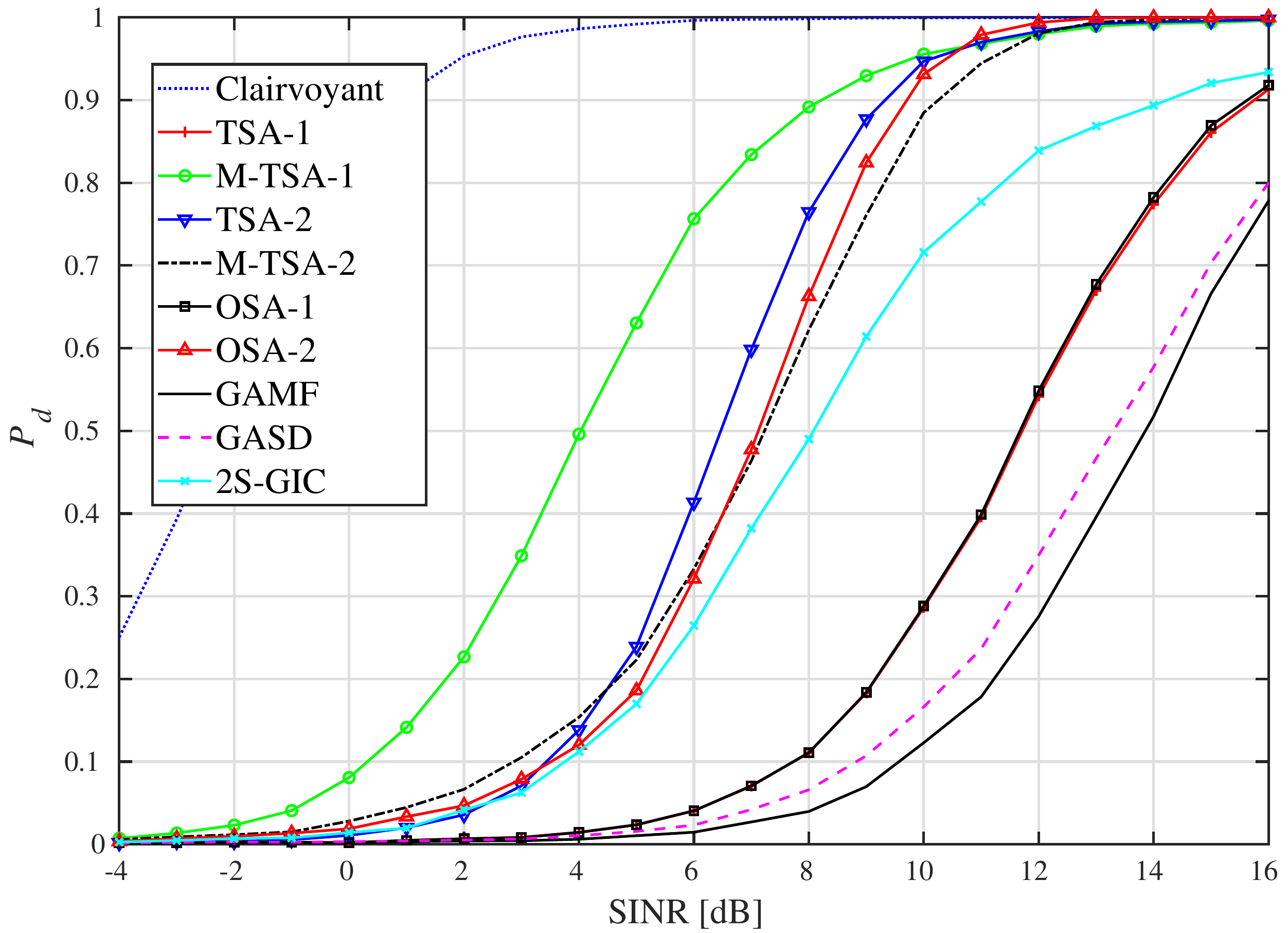}
\label{fig:pd_SINR}
}
\!\!\!\!\!\!\!\!
\subfigure[MVMP versus SINR.]{
\includegraphics[width=0.48\linewidth, height=3.3cm]{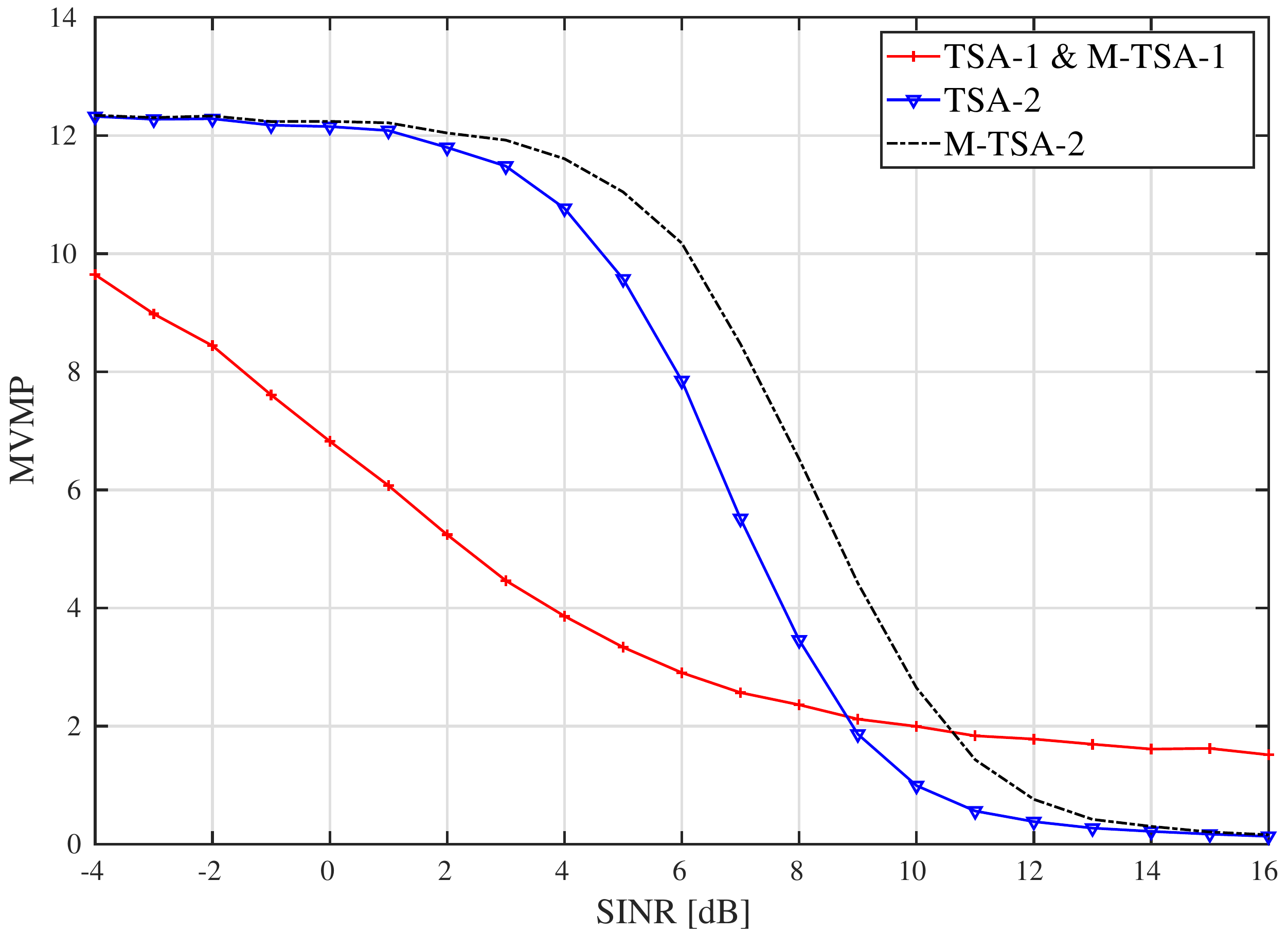}
\label{fig:MVMP_SINR}
}
\subfigure[$P_d$ versus SINR without RCM.]{
\includegraphics[width=0.6\linewidth, height=3.5cm]{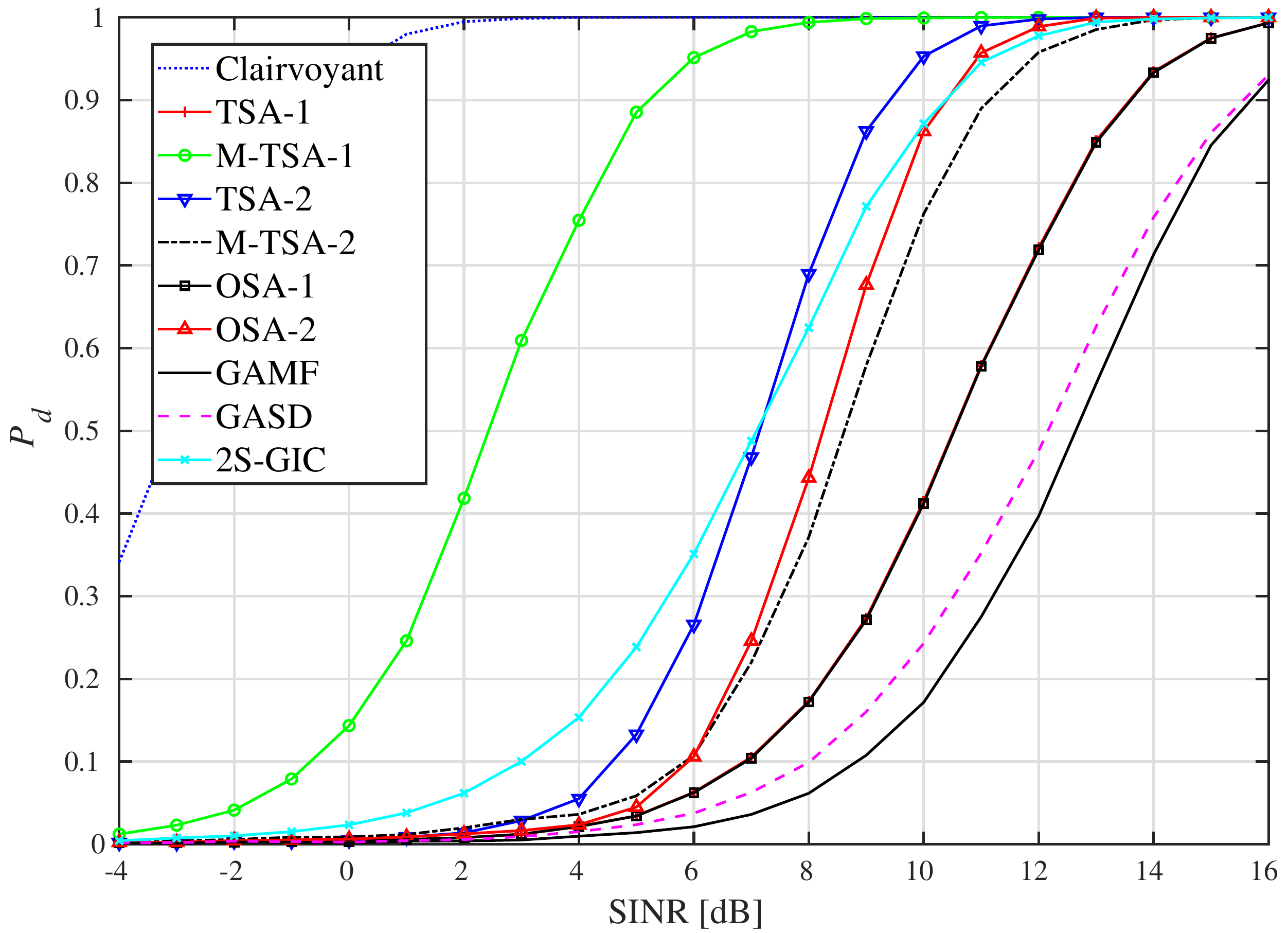}
\label{fig:pd_no_rcm}
}
\caption{Detection and estimation performance.}
\end{figure}

The parameters of the high-resolution radar and maneuvering target are: $f_c=10.0$ GHz, bandwidth 500 MHz,
range resolution 0.3 m, $\mbox{PRT}=1$ ms, $N_p=16$, $N_a=8$ and $v_t=30$ m/s.
In this scenario, the point-like target will occupy more than one range cell during a pulse integration interval.
The related curves of $P_d$ versus SINR (defined as in \cite{DimMovingT}) for $P_{fa} = 10^{-3}$, $K = 12<2N_a$
are shown in Fig. \ref{fig:pd_SINR}. It turns out that
M-TSA-1 overcomes the other competitors with a gain of more than $9$ dB over the GAMF at $P_d>0.5$.
The TSA-2, OSA-2, and M-TSA-2 follow the M-TSA-1 and with $P_d$ values contained in an interval of about $1$ dB.
The 2S-GIC experiences a loss of $4$ dBs with respect to M-TSA-2.
The MVMP curves versus SINR are shown in Fig. \ref{fig:MVMP_SINR}.
Inspection of the figure highlights that for SINR values lower than $10$ dB, architectures based on \eqref{eqn:BIC_2S}
return better estimation results than TSA-2 (rule \eqref{eqn:BIC_1S}) and M-TSA-2 (rule \eqref{eqn:BIC_1S_LCC}).
When $\mbox{SINR}>10$ dB values, \eqref{eqn:BIC_1S} and \eqref{eqn:BIC_1S_LCC} slightly outperform \eqref{eqn:BIC_2S}.
The detection performances when the RCM does not occur are shown in Fig. \ref{fig:pd_no_rcm} for $P_{fa} = 10^{-3}$, $K = 12$, $l = N_p$,
and $h = 0$. The M-TSA-1 still guarantees superior performance over the other detectors.

Finally, we compare the considered architectures from a computational point of view using the usual Landau notation.
As expected, the GAMF is the architecture with the lowest computational load since it does not involve the determinant computation,
data-dependent normalization, and discrete search; its computational load is given by $\mathcal{O}(N_a^{3}+N_a^{2}K+2N_aN_p)$.
The GASD with complexity $\mathcal{O}(N_a^{3}+N_a^{2}(K+2N_p))$ is slightly more time demanding than
the GAMF due to the data-dependent normalization and shares a similar complexity with OSA-1 and TSA-1,
that, in turn, are $\mathcal{O}(N_a^{3}+N_a^{2}K+2N_aN_p+\frac{1}{2}N_p^{2})$.
Proceeding in order of increasing complexity, we obtain that M-TSA-1 and M-TSA-2
are $\mathcal{O}(2N_a^{3}+N_a^{2}(K+2N_p))$ and $\mathcal{O}(\frac{1}{2}N_a^{3}N_p^{2}+\frac{3}{2}N_a^{2}N_p^{2})$, respectively.
The most complex architectures are OSA-2 and TSA-2, which are
$\mathcal{O}(N_a^{3}N_p^{2}+2N_a^{2}N_p^{2}+N_aN_p^{3})$, and 2S-GIC whose complexity is
$\mathcal{O}(N_a^{3}N_p^{2}+\frac{3}{2}N_a^{2}N_p^{2}+\frac{1}{2}N_aN_p^{3})$. As a matter of fact, they require the computation
of $\hat{M}_{l,h}$ and its determinant for each $l\in\Omega_p$ and $h:\, l+h \le N_p$. Summarizing, the analysis singles out the
M-TSA-1 as the architecture that provides an excellent compromise between detection/estimation performance and computational load.

\section{Conclusion}
This letter focused on the adaptive detection of dim maneuvering target in the presence of range migration.
In this context, data containing the returns from two adjacent range cells have been exploited to
conceive six different decision schemes with different computational requirements
that incorporate the BIC rule to estimate the range migration indices.
The performance assessment pointed out that the M-TSA-1 can ensure an excellent trade off between detection performance
and computational cost also for low volumes of training data.
Future research tracks may include the design of architectures accounting for the spillover of target energy or heterogeneous environments.

\bibliographystyle{IEEEtran}
\bibliography{group_bib2}

\end{document}